\documentclass[aps,prb,superscriptaddress,draft,showpacs,intlimits,amsmath,amssymb,floats,floatfix,twocolumn]{revtex4}
\usepackage{bm}
\usepackage[final]{graphicx}
\usepackage{epsfig}
\begin{document}
\title{Band and momentum dependent electron dynamics in superconducting ${\rm
Ba(Fe_{1-x}Co_{x})_2As_2}$ as seen via electronic Raman scattering}
\date{\today}
\author{B. Muschler}
\affiliation{Walther Meissner Institut, Bayerische Akademie der
Wissenschaften, 85748 Garching, Germany}
\author{W. Prestel}
\affiliation{Walther Meissner Institut, Bayerische Akademie der
Wissenschaften, 85748 Garching, Germany}
\author{R. Hackl}
\affiliation{Walther Meissner Institut, Bayerische Akademie der
Wissenschaften, 85748 Garching, Germany}
\author{T.\,P. Devereaux}
\affiliation{Stanford Institute for Materials and Energy
Sciences, SLAC National Accelerator Laboratory,
2575 Sand Hill Road, Menlo Park, CA 94025,~USA}
\author{J.\,G. Analytis}
\affiliation{Stanford Institute for Materials and Energy
Sciences, SLAC National Accelerator Laboratory,
2575 Sand Hill Road, Menlo Park, CA 94025,~USA}
\affiliation{Geballe Laboratory for Advanced Materials \& Dept. of Applied Physics,
Stanford University, Stanford, CA 94305, USA}
\author{Jiun-Haw Chu}
\affiliation{Stanford Institute for Materials and Energy
Sciences, SLAC National Accelerator Laboratory,
2575 Sand Hill Road, Menlo Park, CA 94025,~USA}
\affiliation{Geballe Laboratory for Advanced Materials \& Dept. of Applied Physics,
Stanford University, Stanford, CA 94305, USA}
\author{I.\,R. Fisher}
\affiliation{Stanford Institute for Materials and Energy
Sciences, SLAC National Accelerator Laboratory,
2575 Sand Hill Road, Menlo Park, CA 94025,~USA}
\affiliation{Geballe Laboratory for Advanced Materials \& Dept. of Applied Physics,
Stanford University, Stanford, CA 94305, USA}

\begin{abstract}
We present details of carrier properties in high quality ${\rm Ba(Fe_{1-x}Co_{x})_2As_2}$ single crystals obtained from electronic Raman scattering. The experiments indicate a strong band and momentum anisotropy of the electron dynamics above and below the superconducting transition highlighting the importance of complex band-dependent interactions. The presence of low energy spectral weight deep in the superconducting state suggests a gap with accidental nodes which may be lifted by doping and/or impurity scattering. When combined with other measurements, our observation of band and
momentum dependent carrier dynamics indicate that the iron arsenides may have several competing superconducting ground states.
\end{abstract}

\pacs{78.30.-j, 74.72.-h, 74.20.Mn, 74.25.Gz}

\maketitle
\begin{figure}
  \centering
  \includegraphics[width=7.5cm]{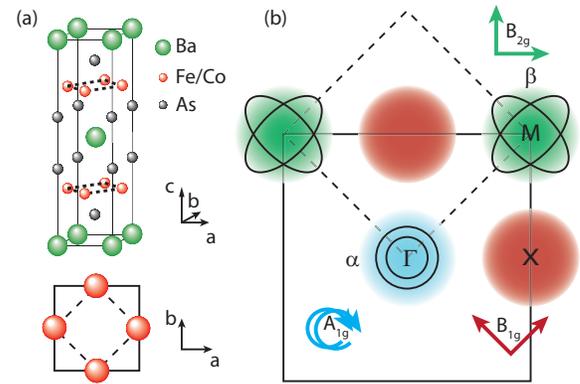}
  \caption[]{(Color online)
  Crystal structure and Raman selection rules for ${\rm
  BaFe_2As_2}$. (a) The relevant cell of the Fe plane is smaller by a factor of
  2 and is rotated by $45^{\circ}$ (dashes) with respect to the crystal
  cell (full line).
  (b) Brillouin zone (BZ) of the unit cell (full line) and first quadrant of the
  Fe plane (dashed line). The
  light polarizations are indicated symbolically with
  respect to the basal plane shown in panel (a). The electronic $A_{1g}$ and $B_{2g}$ spectra project the $\alpha$ and $\beta$ bands, respectively.

  } \label{fig:FeAs}
\end{figure}

The high temperature iron-arsenide (FeAs)
superconductors (Fig.~\ref{fig:FeAs}~a) \cite{Kamihara:2008,Rotter:2008} exhibit a similar
proximity of ground
states \cite{Rotter:2008,Luetkens:2009,Chen:2009,Goko:2009} as some heavy fermion systems and  the
copper-oxygen compounds.
In particular the proximity of the
parent antiferromagnetic to the optimal superconducting phase
suggests that the copper-oxygen compounds and the iron arsenides
may be cousins in the same family. However, the strong metallicity
of the parent phase and the lack of spectral redistribution upon
doping observed by angle-resolved photoemission (ARPES) and x-ray
absorption (XAS) in FeAs argue otherwise
\cite{Lu:2008,Kurmaev:2008}. But perhaps one of the most telling
similarities would be if the iron arsenides had the signature
property of all cuprates -- an energy gap $\Delta_{\bf k}$ having nodes and a
sign change along the Fermi surface  \cite{Tsuei:2000}.

In the iron arsenides $T_c$ and the superconducting gap structure have shown a
remarkable dependence on the material class and on doping. The magnetic penetration depth $\lambda(T)$ clearly indicates nodes in ${\rm LaFePO}$  \cite{Fletcher:2009,Hicks:2009} and  in the ${\rm BaFe_2As_2}$ family
\cite{Gordon:2009,Prozorov:2009}. In SmFeAsO$_{1-x}$F$_{y}$ the small finite gaps derived from $\lambda(T)$ \cite{Malone:2009} indicate either strongly anisotropic gaps or gaps that vary significantly between the different Fermi surface sheets as seen in ARPES in ${\rm Ba_{1-x}K_{x}Fe_2As_2}$ \cite{Evtushinsky:2009} and
${\rm Ba(Fe_{1-x}Co_{x})_2As_2}$ \cite{Terashima:2009}. Small but finite gaps are also derived from recent measurements of the thermal conductivity in ${\rm Ba(Fe_{1-x}Co_{x})_2As_2}$ at very low temperatures \cite{Tanatar:2009}. As an open experimental issue, surface sensitive point contact spectroscopy and ARPES experiments generally observe full gaps \cite{Evtushinsky:2009,Terashima:2009,Wang:2009b,Kondo:2008,Samuely:2009} whereas
bulk sensitive nuclear magnetic resonance (NMR) experiments so far found only indications of gaps with nodes \cite{Grafe:2008,Mukuda:2009}. While on one hand, polar
surfaces (which may distort the nature of the pair state close to
the surface from that of the bulk) may account for the differences
between bulk and surface methods, it is also an intriguing
possibility that different FeAs superconductors may have different
superconducting ground states, such as $s_{\pm}$ or $d-$wave,
which may be selected by small changes in materials
chemistry. 

In this study we use bulk and band sensitive inelastic (Raman) light scattering to gain
insight into electron dynamics and structures of the gaps. Photons are scattered off
of electrons by creating particle-hole pairs across the Fermi
level in the normal state or by breaking Cooper pairs  in the superconducting
state \cite{Devereaux:2007}. By changing incident and scattered
light polarizations, electron dynamics and the superconducting
energy gap can be highlighted in different momentum regions in the
Brillouin zone (BZ) (see Fig.~\ref{fig:FeAs}~b). Since the bands lie at high symmetry points in the BZ $B_{2g}$  ($xy$) predominantly probes the $\beta$ bands, and $B_{1g}$ ($x^2-y^2$) does not couple strongly to any band. While the $A_{1g}$  ($x^2+y^2$) vertex in principle probes both bands the largest contribution represents
interband scattering involving the $\alpha$ bands, modified by charge backflow effects.

The single crystals of electron doped ${\rm Ba(Fe_{1-x}Co_{x})_2As_2}$ with
$x=0.061(2)$ and $x=0.085(2)$ were synthesized using a self-flux technique and have
been characterized elsewhere \cite{Chu:2009}. The cobalt
concentration was determined by microprobe analysis. At 6.1\%  and 8.5\% doping
the structural and magnetic transitions have been suppressed below
the superconducting transition at $T_c = 24$~K and 22~K, respectively, with $\Delta T_c <1~K$. The
resistivity varies essentially linearly between $T_c$ and 300~K.

\begin{figure}
  \centering
  \includegraphics[width=5.5cm]{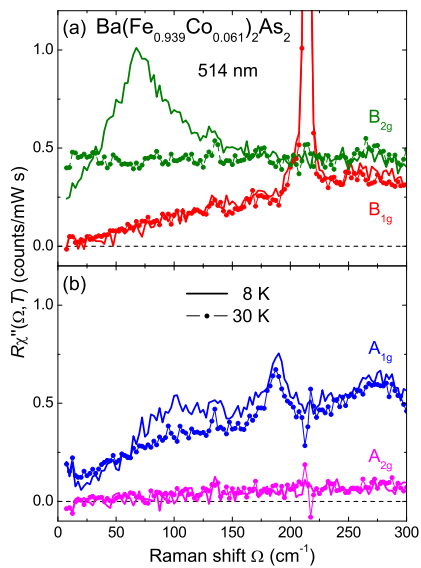}
  \caption[]{(Color online)
  Symmetry-dependent Raman response $R\chi^{\prime\prime}(\Omega,T)$ of ${\rm
  Ba(Fe_{0.939}Co_{0.061})_2As_2}$.
  (a) In $B_{1g}$ symmetry there is a phonon at 214~cm$^{-1}$ from Fe vibrations.
  (b) In $A_{1g}$ there is a small increase towards $\Omega \rightarrow 0$
  since it is always measured with parallel
  polarizations where the laser is less efficiently suppressed.
  The analysis demonstrates that the $A_{2g}$
  signal can safely be neglected.\cite{A2g}
  } \label{fig:raw}
\end{figure}

In Fig.~\ref{fig:raw} we plot the spectra $R\chi^{\prime\prime}$
measured on ${\rm Ba(Fe_{0.939}Co_{0.061})_2As_2}$  for the 4
distinct in-plane symmetries which are linear combinations of the spectra measured at the principal polarizations. We show spectra at
$8$~K for the superconducting state and at 30~K for the
normal state. There is a strong dependence on symmetry in either
state, indicated by the differences in the overall spectral lineshape and
intensity, but also by the different temperature
dependences.\cite{new_ms}

The $B_{2g}$ spectra (Fig.~\ref{fig:raw} a) are strikingly
different from those in the $A_{1g}$ and $B_{1g}$ symmetries. In the normal state,
the flat electronic continuum is similar to that in the cuprates \cite{Devereaux:2007} and changes strongly with temperature.\cite{new_ms} This indicates
that the electrons in the $\beta$ bands scatter dynamically from excitations.
In contrast, the $A_{1g}$,  $B_{1g}$, and $A_{2g}$  symmetries (Fig.~\ref{fig:raw}) yield
suppressed practically temperature independent spectra.\cite{new_ms}
This strong polarization
dependence implies that the charge carrier relaxation
on the $\alpha$ bands ($A_{1g}$) is fundamentally
different from that on the $\beta$ bands ($B_{2g}$) and
indicates strongly anisotropic and band dependent electron interactions.

\begin{figure*}
  \centering
  \includegraphics [width=14cm]{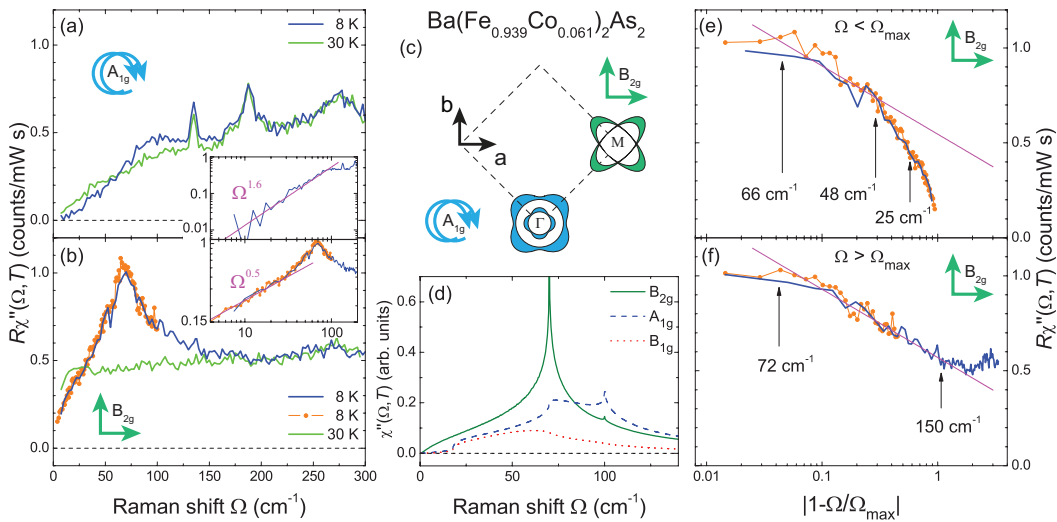}
  \caption[]{(Color online)
  Raman response $R\chi^{\prime\prime}(\Omega,T)$ of ${\rm
    Ba(Fe_{0.939}Co_{0.061})_2As_2}$ in (a) $A_{1g}$ and (b) $B_{2g}$ polarizations. Spectra plotted with full lines are measured with a resolution of 5.0~cm$^{-1}$. For further clarifying the spectral shape at low energy and around the maximum the superconducting $B_{2g}$ spectra were also measured with a resolution of 3.6~cm$^{-1}$  (orange points in b, e, f).
    The insets indicate power-law behavior for both symmetries. The finite spectral intensity at low energies
    supports  gap nodes. (c) shows the gap forms used for the Raman spectra calculated in (d). The $A_{1g}$ and $B_{1g}$ spectra are multiplied by 0.5 and 10, respectively. (e) and (f) show the $B_{2g}$ spectra above and below the peak maximum on a logarithmic
    scale to highlight the divergence around $\Omega_{\rm max}=69$~cm$^{-1}$.
  } \label{fig:slope}
\end{figure*}

These polarization dependences carry through into the
superconducting state, indicating a marked difference of light
scattering from Cooper pairs on the $\alpha$ and $\beta$ bands (Fig.~\ref{fig:raw}). While for $B_{1g}$ symmetry
there is no difference between the normal and superconducting
state, for $B_{2g}$ a strong peak at around 70 cm$^{-1}$ and a
suppression of spectral weight below 30 cm$^{-1}$ develops below
$T_c$. Here, the peak intensity and its resolution-limited
sharpness (see Fig.~\ref{fig:slope} e and f) emphasize the high
purity and order of the sample used. The $A_{1g}$ spectra below
$T_c$ (Fig.~\ref{fig:raw}~b) reach a broad maximum
close to 100~cm$^{-1}$. This higher energy scale indicates a
larger gap amplitude on the $\alpha$ bands than on the $\beta$
bands. The maximum at 100~cm$^{-1}$ or
$6\,k_BT_c$ is consistent with the photoemission results
\cite{Evtushinsky:2009,Terashima:2009}.

The redistribution of spectral weight and the peak structures are
reminiscent of the superconductivity-induced features in the A15
compounds and in overdoped cuprates
\cite{Devereaux:2007}. However,
unlike the A15s, and more like the cuprates, the finite intensity
observed down to very small Raman shifts is a clear indication of
vanishingly small gaps,
where Cooper pairs can be broken with arbitrarily low energies.

In Fig.~\ref{fig:slope} a and b we show the $A_{1g}$ and $B_{2g}$ spectra in greater detail.
Both do not
show a clear activation threshold but finite intensity down to
arbitrarily small Raman shifts, favoring the presence of nodes
rather than full gaps on at least some of the bands. Below $80$~cm$^{-1}$ the $A_{1g}$ intensity
varies faster than linear following
$\Omega^{1.6}$ (inset of Fig.~\ref{fig:slope} a). If there is a
threshold it must be smaller than $30$~cm$^{-1}$. This would
translate into a minimal gap $\Delta_{\rm min} \le 2$~meV or
$0.9\,k_BT_c$ much smaller than that observed by ARPES
\cite{Evtushinsky:2009}. Importantly, the $B_{2g}$ spectrum
(Fig.~\ref{fig:slope} b) varies approximately as $\sqrt{\Omega}$ with a
potential threshold below $5$~cm$^{-1}$ suggesting the presence of accidental gap nodes.

The different power laws for $A_{1g}$ and $B_{2g}$
indicate a sensitivity of the pairing gap to the Fermi surface
location, area, and perhaps, geometry. The low-frequency
power laws of the Raman response follow from the
density of states (DOS) unless the nodes of the Raman vertices
happen to be aligned with the nodes of the gap, as in the case of the $B_{1g}$ channel in the cuprates
\cite{Devereaux:2007}, imparting a higher power law than that
expected from the DOS alone. Since no nodes of the vertex are required by symmetry on the $\beta$ bands the $\sqrt{\Omega}$ behavior in the
$B_{2g}$ channel argues strongly for a superconducting gap
vanishing quadratically with momentum near the nodes. This is the case for
an $s$-wave pair state having accidental nodes (see Fig.~\ref{fig:slope} (c) and Ref.~\cite{new_ms}).

\begin{figure}
  \centering
  \includegraphics [width=6cm]{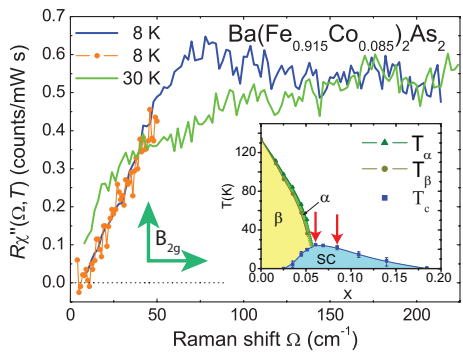}
  \caption[]{(Color online)
  $B_{2g}$ Raman scattering response $R\chi^{\prime\prime}(\Omega,T)$ of ${\rm
  Ba(Fe_{0.915}Co_{0.085})_2As_2}$. Spectra plotted with full lines (orange points) are measured with a resolution of 5.0~cm$^{-1}$ (3.6~cm$^{-1}$). The inset shows the phase diagram \cite{Chu:2009} with the two samples studied here indicated by arrows. Instead of the sublinear energy dependence at 6.1\% (Fig.\ref{fig:slope} (b)) a finite gap of approximately 10~cm$^{-1}$ is clearly resolved here. The logarithmic singularity is replaced by a broad maximum. Both the low- and the high-energy parts of the spectra follow naturally from the influence of impurities.\cite{Devereaux:1995}
  } \label{fig:raw085}
\end{figure}

On the $\alpha$ bands there are no nodes of the Raman vertices either. Therefore the
observed $\Omega^{1.6}$ dependence of the intensity for the $A_{1g}$ response
cannot originate from an interplay between the nodal structure of the gap and the
vertices; rather it most likely reflects a threshold broadened by incoherent scattering. These considerations show that there is no universal
superconducting energy gap. Rather, there is a substantial
variation for the different sheets of the Fermi surface as well as
a strong momentum dependence on the individual sheets. Hence, bulk spectroscopic methods projecting individual sheets of the Fermi surface such as Raman scattering add information relevant for the understanding of the superconductivity in the pnictides which is hardly accessible by probes which yield the integral response of all Fermi surface sheets. 

The multi-gap behavior shown in Fig.~\ref{fig:slope}~c yields the best agreement with theoretical
predictions \cite{Boyd:2009} as plotted in
Fig.~\ref{fig:slope}~(d). In particular the following features
can be reproduced: (i) the superlinear variation and the broad
maximum in $A_{1g}$ symmetry indicating a strong variation of the
gaps on the $\alpha$ sheets, (ii) the
vanishingly small contribution in $B_{1g}$ symmetry due to
matrix-element effects, and (iii) the entire $B_{2g}$ spectra
including the sublinear variation below 30~cm$^{-1}$ due to
accidental nodes on the $\beta$ sheets and the logarithmic
variation around the cusp-like maximum. When we plot the spectra
as a function of $\log(|\Omega-\Omega_{\rm max}|/\Omega_{\rm
max})$ we find indeed a universal linear variation on either side
of the maximum at $\Omega_{\rm max} = 69$~cm$^{-1}$ which extends
over a decade and half a decade on the high- and the low-energy
sides, respectively (Fig.~\ref{fig:slope} e and f). The divergence
is cut only by the resolution of the spectrometer.

The sharp structures observed here open up an opportunity to study the effect of impurities. According to our normal state results but also to transport \cite{Chu:2009} the residual scattering rate is of the same order of magnitude as the gap.\cite{new_ms} Disorder is expected to cut the singularity and to open up a finite gap around the nodes  \cite{Devereaux:1995,Mishra:2009}. To investigate this issue, we have measured ${\rm Ba(Fe_{0.915}Co_{0.085})_2As_2}$ with a slightly reduced $T_c$. As shown in Fig.~\ref{fig:raw085}, the $B_{2g}$ spectra indicate a finite threshold and a reduced peak height at around 70~cm$^{-1}$. These results demonstrate that the energy gap can be dramatically affected by sample properties. It is an open question of how the gap and transport dynamics on different portions of the bands and Brillouin zone are modified by doping and disorder. The effects of changes on the pairing interaction and the scattering rates need to be disentangled, which opens a area of further investigation  and suggests that a number of apparently contradictory experiments may be reconciled by considering sample quality and distinguishing bulk versus surface sensitive measurements.

The observed power laws arguing for accidental nodes on the
$\beta$ sheets and near-nodes on the $\alpha$ sheets are
inconsistent with ARPES \cite{Evtushinsky:2009,Terashima:2009} but consistent with penetration depth and thermal conductivity studies \cite{Hicks:2009,Prozorov:2009,Tanatar:2009,Mishra:2009a}. A possible distinction may be made when one considers probing the surface versus probing the bulk as Raman scattering
does. Yet, another intriguing possibility is that there may be
competing superconducting instabilities that can be triggered by
small changes in carrier concentrations. This appears to be the case
in several numerical simulations of spin fluctuation models using
both fluctuation exchange in random phase approximation (RPA)
\cite{Kuroki:2008,Graser:2009} and numerical functional
renormalization group \cite{Wang:2009a} calculations. These
calculations find that $s-$wave ($A_{1g}$) and $d$-wave ($B_{2g}$)
instabilities can occur in multi-band models of the Fe-pnictides,
which lie relatively close to each other. Moreover, the $s$-wave
pair state is found to have substantial anisotropy. Finally, since
our samples are still very close to a spin density wave (SDW) state,
it is plausible that the order parameters for superconductivity
and SDW magnetism couple. The resulting apparent gap anisotropy
might not exist without SDW order or fluctuations. At the doping level studied we could not find indications of collective modes such as proposed for the case of an $s_{\pm}$ state \cite{Chubukov:2009}.

Thus, Raman measurements show that for ${\rm
Ba(Fe_{1-x}Co_{x})_2As_2}$ different
nodal or near-nodal behavior occurs on the $\alpha$ and $\beta$ Fermi surface
sheets. In either case impurity scattering is of crucial importance when analyzing the superconducting gap in the pnictides. The polarization dependence of the data both in the
superconducting and normal states argues for anisotropic charge
dynamics on the different Fermi surface sheets - a signature of
strongly momentum dependent particle interaction. Nesting
properties between the $\alpha$ and $\beta$ sheets rather than
electron-phonon coupling would provide a strongly enhanced dynamic
interaction between electrons from nearest neighbor Fe orbitals.
In this way the iron pnictides would be indeed close relatives of
the cuprates, the ruthenates, some heavy-fermion compounds or even
$^3$He being at the brink of stability of various ground states
\cite{Monthoux:2007}.

\textbf{Acknowledgements:}   We acknowledge support by the DFG under grant number HA 2071/3 via
  			 Research Unit FOR538. The work at SLAC and Stanford University is supported
  			 by by the Department of Energy, Office of Basic Energy Sciences under
  			 contract DE-AC02-76SF00515. R.H. and T.P.D. gratefully acknowledge support by the KITP and by the
             Bavarian-Californian Technology Center (BaCaTeC).


\begin{thebibliography}{10}
\bibitem{new_ms}
{{Details will be presented in an upcoming publication.}}
\bibitem{A2g}
{{The spin-chirality operator, crystal-field excitations or
  interorbital transitions can have $A_{2g}$ symmetry. $A_{2g}$ contributions
  are also expected if the light-scattering process is not invariant under time
  reversion, for instance if the photon energy is equal to an interband
  transition}}.
\end{thebibliography}
\end{document}